\documentclass[
a4paper,
twocolumn,
showpacs,
superscriptaddress,
floatfix,
nofootinbib,
prx,
longbibliography,
]{revtex4-2}
\usepackage{hyperref}
\hypersetup{
    colorlinks=true,
    linkcolor=blue,
    filecolor=magenta,      
    urlcolor=blue,
    pdftitle={Overleaf Example},
    pdfpagemode=FullScreen,
    }

\usepackage{array}
\usepackage{graphicx}
\usepackage{amssymb,amsmath}
\usepackage[usenames]{color}
\usepackage{amssymb}
\usepackage{wasysym}
\usepackage{soul}
\usepackage{times}
\usepackage{mathrsfs}
\usepackage{hyperref}
\usepackage{booktabs}
\usepackage{chngcntr}
\hypersetup{
  colorlinks=true,        
  linkcolor=blue,         
  citecolor=cyan,         
}

\usepackage{float}
\usepackage{tensor}

\definecolor{orange}{rgb}{1,0.5,0}

\def\prd{{\it Phys. Rev.} D~}

\def\mnras{{\it MNRAS~}}

\def\lesssim{\mathrel{\hbox{\rlap{\hbox{\lower4pt\hbox{$\sim$}}}\hbox{$<$}}}}
\def\gtrsim{\mathrel{\hbox{\rlap{\hbox{\lower4pt\hbox{$\sim$}}}\hbox{$>$}}}}
\def\alt{\mathrel{\hbox{\rlap{\hbox{\lower4pt\hbox{$\sim$}}}\hbox{$<$}}}}
\def\agt{\mathrel{\hbox{\rlap{\hbox{\lower4pt\hbox{$\sim$}}}\hbox{$>$}}}}

\usepackage{mathtools}

\newenvironment{cititemize2}
{\begin{list}{$\bullet$}
        {\setlength{\topsep}{0pt}
         \setlength{\itemsep}{0pt}
         \setlength{\parsep}{0.25\parsep}
         \settowidth{\labelwidth}{$\bullet$}
         \setlength{\leftmargin}{1em}
}
}
{\end{list}}

\def\gta{\ifmmode {\mathbin{\lower 3pt\hbox   
    {$\,\rlap{\raise 5pt\hbox{$\char'076$}}\mathchar"7218\,$}}}
    \else {${\mathbin{\lower 3pt\hbox
    {$\rlap{\raise 5pt\hbox{$\char'076$}}\mathchar"7218\,$}}}
    $}\fi}
\def\lta{\ifmmode {\,\mathbin{\lower 3pt\hbox   
    {$\,\rlap{\raise 5pt\hbox{$\char'074$}}\mathchar"7218\,$}}}
    \else {${\mathbin{\lower 3pt\hbox
    {$\rlap{\raise 5pt\hbox{$\char'074$}}\mathchar"7218\,$}}}
    $}\fi}

\newcommand{\beq}{\begin{equation}}
\newcommand{\eeq}{\end{equation}}
\newcommand{\bea}{\begin{eqnarray}}
\newcommand{\eea}{\end{eqnarray}}

\renewcommand{\BibitemShut}[1]{}

\graphicspath{
{figures/}
{figures/Burgers1D/}
{figures/Burgers2D/}
{figures/Burgers2D_novisc/}
{figures/Burgers2D_coupled/}
{figures/SWE_Linear/}
{figures/SWE_Nonlinear/}
{figures/SWE_Nonlinear_extra/}
{figures/Wave1D/}
{figures/Wave2D/}
{figures/Wave2D_nonconst_coeff/}
{figures/Burgers2D_FNO/}
}

\DeclareGraphicsExtensions{.pdf}

\begin{document}
\title{Applications of physics informed neural operators}

\author{Shawn G. Rosofsky}
\affiliation{Data Science and Learning Division, Argonne National Laboratory, Lemont, Illinois 60439, USA}
\affiliation{Department of Physics, University of Illinois at Urbana-Champaign, Urbana, Illinois 61801, USA}
\affiliation{NCSA, University of Illinois at Urbana-Champaign, Urbana, Illinois 61801, USA}
\author{Hani Al Majed}
\affiliation{Data Science and Learning Division, Argonne National Laboratory, Lemont, Illinois 60439, USA}
\affiliation{NCSA, University of Illinois at Urbana-Champaign, Urbana, Illinois 61801, USA}
\affiliation{Department of Electrical and Computer Engineering, University of Illinois at Urbana-Champaign, Urbana, Illinois 61801, USA}
\author{E. A. Huerta}
\affiliation{Data Science and Learning Division, Argonne National Laboratory, Lemont, Illinois 60439, USA}
\affiliation{Department of Physics, University of Illinois at Urbana-Champaign, Urbana, Illinois 61801, USA}
\affiliation{Department of Computer Science, The University of Chicago, Chicago, Illinois 60637, USA}

\begin{abstract}
\noindent We present an end-to-end framework to learn partial differential equations that brings together initial data production, selection of boundary conditions, and the use of physics-informed neural operators to solve partial differential equations that are ubiquitous in the study and modeling of physics phenomena. We first demonstrate that our methods reproduce the accuracy and performance of other neural operators published elsewhere in the literature to learn the 1D wave equation and the 1D Burgers equation. Thereafter, we apply our physics-informed neural operators to learn new types of equations, including the 2D Burgers equation in the scalar, inviscid and vector types. Finally, we show that our approach is also applicable to learn the physics of the 2D linear and nonlinear shallow water equations, which involve three coupled partial differential equations. We release our artificial intelligence surrogates and scientific software to produce initial data and boundary conditions to study a broad range of physically motivated scenarios. We provide the \href{https://github.com/shawnrosofsky/PINO_Applications/tree/main}{source code}, an interactive \href{https://shawnrosofsky.github.io/PINO_Applications/}{website} to visualize the predictions of our physics informed neural operators, and a tutorial for their use at the \href{https://www.dlhub.org}{Data and Learning Hub for Science}.\end{abstract} 

\maketitle

\section{Introduction}
\label{sec:intro}
The description of physical 
systems has a common 
set of elements, namely: the use of 
fields (electromagnetic, gravitational, etc.) 
on a given spacetime manifold, a geometrical 
interpretation of these fields in terms of the 
spacetime manifold, partial differential equations 
(PDEs) on these fields that describe the change 
of a system over spacetime, and an initial 
value formulation of these equations 
with suitable 
boundary conditions~\cite{Geroch:1996kg}. 
Given that the evolution of physical fields 
over spacetime may be naturally expressed 
in terms of PDEs, a 
plethora of numerical methods have been developed 
to accurately and rapidly solve these 
class of equations~\cite{nu_re}.

In time, and even with the advent of 
extreme scale computing, some physical 
systems have become increasingly difficult 
to model, e.g., multi-scale and 
multi-physics systems that combine 
disparate time and spatial scales, and 
which demand the use of subgrid-scale 
precision to accurately resolve the 
evolution of physical fields. This is a well 
known problem in multiple disciplines, 
including general relativistic 
simulations~\cite{2020Symm...12.1249R,2021arXiv211114858R,Summ_Fou}, 
weather forecasting~\cite{Weather_Forecasting}, 
\textit{ab initio} density functional theory simulations~\cite{DFT_sims}, 
among many other computational grand challenges.

The realization that large scale computing 
resources are finite and will continue to be 
oversubscribed~\cite{HPCUSE,2020arXiv201209303G}, has 
impelled scientists to explore novel approaches to 
address computational bottlenecks in scientific 
software~\cite{2020arXiv200308394H}. Some approaches include rewriting modules 
of software stacks to leverage GPUs, leading to 
significant 
speedups~\cite{sims_gpus,gpu_proteins,2019PhRvD..99h4026W}. Other contemporary approaches have 
harnessed advances in machine learning to 
accelerate specific computations in 
software modules~\cite{bambi:2012MNRAS}, while 
others have developed 
entirely new solutions by combining 
GPU-accelerated computing and novel signal 
processing tools that have at their core 
machine learning or artificial intelligence 
applications~\cite{2020PhRvD.101h4024R, huerta_book,cuoco_review,huerta_nat_ast,Nat_Rev_2019_Huerta,khan_huerta_zheng_forecast,Chaturvedi:2022suc}.

The creation of AI surrogates aims not only 
to enhance the science reach of advanced 
computing facilities. Most importantly, 
this emergent area of research aims to combine 
AI and extreme scale computing to enable 
research that would 
otherwise remain unfeasible with traditional 
approaches. Furthermore, 
AI surrogates aim to capture known knowledge, and 
first principles to stir AI learning in the 
right direction, and then refine AI surrogates 
performance and predictions by using experimental 
scientific data. In time, it is expected that 
by exposing AI to detailed simulations, 
first principles and experimental data, 
AI surrogates will capture the nonlinear behaviour 
of experimental phenomena and guide the 
planning, automation and execution of new 
experiments, leading to breakthroughs in science 
and engineering.

In this article we contribute to the construction of 
AI surrogates by demonstrating their application 
to solve a number of PDEs that are ubiquitous in 
physics, and which have not been presented before 
in the literature. 
{
Specifically, we have three levels of applicability 
of these new problems.
First, we performed sanity checks with simple
PDEs to verify that our model behaves as expected.
Second, we tested new numerically challenging cases
to assess the applicability of PINOs under such conditions,
which include non-constant coefficients, coupled PDEs,
and shocks.  Third, we applied PINOs to coastal and tsunami
modeling with the 2D linear and nonlinear shallow water equations.
}
Beyond these
original contributions, we also provide an end-to-end 
framework that unifies initial data production, 
construction of boundary conditions and their 
use to train, validate and test the performance 
and reliability of AI surrogates. These activities 
aim to create FAIR (findable, accessible, interoperable 
and reusable) and AI-ready datasets 
and models~\cite{fair_article,fair_hep}. 

In the following 
sections we introduce  
key concepts and ideas that will facilitate the 
understanding and use of physics-inspired neural 
operators (PINOs) to solve PDEs \cite{PINO_li_anima}. This article 
is organized 
as follows. In Section~\ref{sec:Modeling} we introduce 
the AI tools we use to learn the physics 
of PDEs. We describe our methods and approaches to create 
PINOs in Section~\ref{sec:Methods}. 
We summarize the PDEs we consider in this study 
in Section~\ref{sec:Tests}, and present a direct 
comparison between numerical solutions of these PDEs and 
predictions from our PINOs in Section~\ref{sec:results}. 
We describe future 
directions of work in Section~\ref{sec:conclusions}.

\begin{figure*}[htb]
\includegraphics[width=\textwidth]{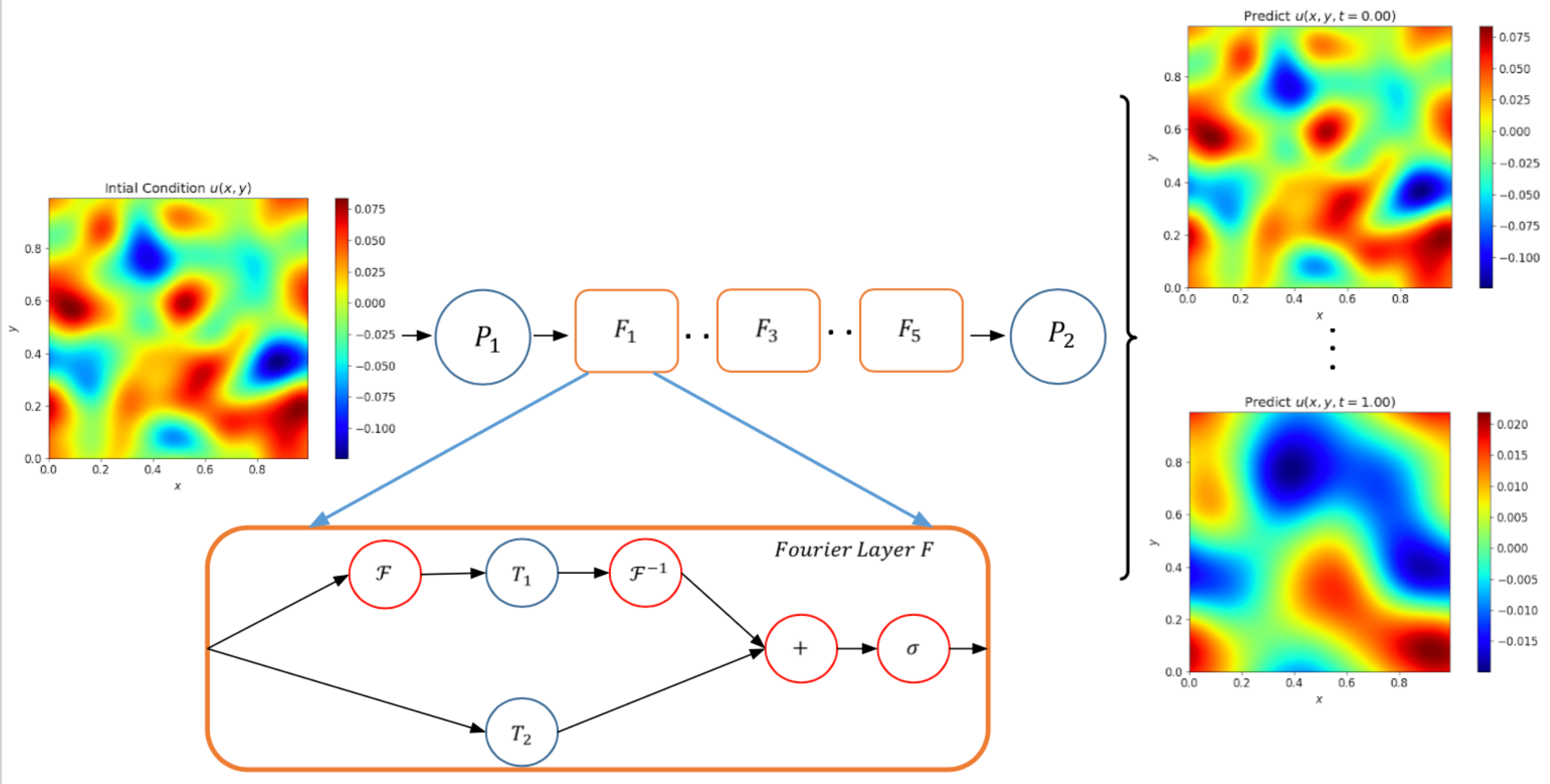}
\centering
\caption{\textbf{Neural network architecture} The top 
panel shows the architecture of our neural operators, 
whereas the inset in the bottom panel shows the structure 
of one of the five Fourier layers, \(\{F_1, ..., F_5\}\), 
we use in our PINOs. 
The top left panel shows the data we feed into our models, labelled as ``Initial Condition \(u(x,y)\)''. This input 
data are initially lifted into a higher dimension 
representation by the neural network, \(P_1\). Thereafter, 
we apply a series of updates that consist of non-local 
integral operators and nonlinear activation 
functions, \(\sigma\), shown in the inset labelled as 
``\textit{Fourier Layer F}''. Eventually, the neural 
network \(P_2\) 
projects back the updates, producing the output shown on the right panels, which 
describe the time 
evolution of the 
system. In the inset, 
\(T_1\) represents a linear transform, \(T_2\) a local 
linear transform, \(\sigma\) a nonlinear activation 
function. \(\mathcal{F}\) and \(\mathcal{F}^{-1}\) 
stand for Fourier transform and inverse 
Fourier transform, respectively.}
\label{fig:architecture}
\end{figure*}

\section{Modeling physics with artificial neural networks}
\label{sec:Modeling}

\subsection{Physics informed neural networks}
\label{sec:PINN}
Physics informed neural networks (PINNs) provide a method of using known physical laws to predict 
the results of various physical systems at high accuracy~\cite{raissi2017physicsI,raissi2017physicsII,raissi2019physics,Pang2019fPINNs,lu2021deepxde}.  These methods estimate 
the results for a given physical system consisting of a 
PDE, initial conditions (ICs), and boundary conditions (BCs) 
by minimizing constraints in the loss function.  PINNs utilize 
the automatic differentiation of deep learning frameworks 
to compute the derivatives of the PDE to compute the 
residual error.  Despite the success of PINNs, their inability 
to produce results for different initial conditions prevents 
them from being useful surrogate models.

\subsection{Neural operators}
\label{sec:NO}

Neural operators use neural networks to learn operators 
rather than single physical systems.  In our case, PDEs 
are the operator these networks try to learn.  
{
Specifically, we provide neural operators with input fields $\mathcal{A}$
that are composed of coordinates and relevant data such as coefficient fields, 
ICs, and BCs.  Neural  operators then output the solutions of that operator 
at those coordinates which we will denote as $\mathcal{U}$.  We can mathematically represent
the neural operator $\mathcal{G}_\theta$ as a mapping between the input fields $\mathcal{A}$ 
and the output fields $\mathcal{U}$ as
\begin{align}
    \label{eq:neural_operator}
    \mathcal{G}_\theta: \mathcal{A} \rightarrow \mathcal{U}
\end{align}
\noindent where $\theta$ are the weights of the neural operator~\cite{kovachki2021neuraloperator}.
}

There are various types of 
neural operators that have been studied in recent 
works such as DeepONets, physics informed DeepONets, 
low-rank neural operators (LNO), 
graph neural operators (GNO),  multipole graph neural operators 
(MGNO), Fourier neural operators (FNO), and physics 
informed neural operators~\cite{lu2021learning,wang2021learning, Li2020GraphNO,Li2020MGNO,Li2021FourierNO,kovachki2021neuraloperator,PINO_li_anima}.  These networks have 
all illustrated their ability to reproduce the results 
of operators much faster than computing with the 
operator directly.  
{
For a more detailed look at neural operators, we refer the 
reader to~\cite{kovachki2021neuraloperator}, which provides
rigorous definitions of neural operators and examines
a large number of different neural operators.
}
\subsection{Physics informed neural operators (PINOs)}
\label{sec:PINO}
PINOs are a variation of neural operators that 
incorporate knowledge of physical laws into their 
loss functions~\cite{PINO_li_anima}.  PINOs have been 
shown to reproduce the results of operators with 
remarkable accuracy. They employ the FNO architecture 
which applies a fast Fourier transform (FFT) to the data 
and applies its fully connected layers in Fourier space 
before performing an inverse FFT back to 
real space~\cite{Li2021FourierNO}.  Moreover, this 
architecture has demonstrated the ability to 
perform zero-shot super-resolution, 
predicting on higher resolution data having only seen 
low resolution 
data~\cite{Li2021FourierNO,kovachki2021neuraloperator}.
Figure~\ref{fig:architecture} illustrates the FNO 
architecture that we use in this study.

PINOs improve upon the FNO architecture by adding 
physics information such as PDEs, ICs, BCs, and 
other conservation laws.  
By including the violation of such laws into the loss function, 
the network can learn these laws in addition to the data.  
Rather than using automatic differentiation, these networks 
use Fourier derivatives to compute the derivatives for 
the PDE constraints as automatic differentiation 
is very memory intensive 
for this type of architecture. This physics knowledge 
enables the network to learn operators faster and with less training data.

\section{Methods}
\label{sec:Methods}

Here we describe the approach we followed to design, 
train, validate and test our PINOs, and then how we 
quantified their accuracy by comparing our 
predictions with actual numerical solutions of 
PDEs we consider in this study.

\subsection{Initial data}
The first step in this process was to generate initial 
data that matches the boundary conditions of the problem. 
This was done using the Gaussian random fields (GRF) 
method in a similar fashion to~\cite{PINO_li_anima} where 
the kernel was transformed into Fourier space to 
match our periodic boundary conditions. We employed a 
general Matern kernel, but discovered that the PINOs 
performed better when the smoothness parameter was set 
to infinity. In this limit, this Matern kernel becomes the 
radial basis function kernel (RBF) as used 
in~\cite{lu2021learning} defined as $k_{l}\left(x_{1}, x_{2}\right)=\exp \left(-\left\|x_{1}-x_{2}\right\|^{2} / 2 l^{2}\right)$, where $l$ is the length scale of typical 
spatial deviations in the data.  For this work, we set $l=0.1$ for all cases to provide features at our desired spatial scale except when explicitly stated otherwise.  A number of these random fields were produced for each of the test problems described Section \ref{sec:Tests}.

{
During this work, we found that the magnitude of the training data 
affected the results of the training even in linear problems.
In particular, the models seem to have difficulty with 
higher magnitude initial data, especially when the data
had a magnitude greater than 1.  We found that by reducing
the magnitude of the initial data, we could improve our
results.  We believe this is attributed to the highly nonlinear
nature of the neural network model.  Other neural network models
also experience similar effects and therefore normalize their
data to improve performance.  Thus, we were careful in selecting the magnitude 
of our input fields during this study.
}

\subsection{Boundary conditions} 
For our boundary conditions, we used periodic boundary 
conditions in all cases.  This boundary condition is good 
for problems with significant symmetry over long length 
scales or for cases where the boundary is sufficiently far 
from the region of interest.  We are particularly focused 
on the latter case where the boundary condition does 
not significantly impact the problem. Moreover, the neural 
network architecture implements periodic boundary conditions 
by default via FFTs.  This periodicity of a certain 
dimension can be removed by zero padding for said dimension 
before feeding it to the neural network. We employed this 
padding in the time dimension with a length of $l_{pad}=5$ 
for all cases except the 1D wave equation, which is 
periodic in time for a time interval of $t=1$.

{
Although we do not do so in this work, one could also
use this technique for PDEs with non-periodic BCs
by zero padding the spatial dimensions
of the data. 
We could then add an additional loss term for the BC 
to the loss in Equation~\ref{eq:loss} to describe our
desired BC.  We describe how to model similar terms in more detail
in Section~\ref{subsec:training_approach}.
}

\subsection{Training data generation}
To generate training data, we took the random IC fields 
that we had previously generated and evolved them in 
space and time. Specifically, we evolve each of these 
equations in time with RK4 time stepping starting at $t=0$ 
until $t=1$ with the timestep $\delta t$ varying depending 
on the problem.  To compute spatial derivatives, 
we employed a finite difference method (FDM) with 4th 
order central difference for most cases.  The lone 
exception was the inviscid Burgers equation in 2D which 
required a shock capturing method.  Therefore, we employed 
a finite volume method (FVM) with local Lax-Friedrichs 
(LLF) fluxes and MP5 reconstruction.

\subsection{Training approach}
\label{subsec:training_approach}
We set up the problem as follows. We are given some space 
and time coordinates as well as the initial conditions 
at those coordinates.  Our objective is to compute the 
solution at each given space and time coordinate from 
the initial data.  To train the network to reproduce 
the simulated results, we employ multiple losses to ensure 
the network properly reproduces the correct loss.  These 
losses are the data loss $\mathcal{L}_{data}$, the physics 
loss $\mathcal{L}_{phys}$, and the IC loss $\mathcal{L}_{IC}$. 
We do not include a loss term for the BC here because the 
FNO architecture of the PINO assures the desired periodic 
BCs as long as those dimensions are not padded.

$\mathcal{L}_{data}$ attempts to fit the model predictions 
directly to the training data.  This loss is computed 
via the relative mean squared error (MSE) between the 
training data and the network outputs.  This relative 
MSE is computed by dividing the MSE of the predictions 
by the norm of the true values.  For cases with multiple 
equations where the outputs vary in magnitude such as the 
linear and nonlinear shallow water equations, care must be
taken to ensure that each of those outputs contributes 
equally to $\mathcal{L}_{data}$.  Therefore, we compute 
the relative MSE for each of the output fields separately 
before combining them together.

$\mathcal{L}_{phys}$ ensures that the PINO predictions 
obey known physical laws such as PDEs or more general 
conservation laws.  We define the loss as the MSE 
between the violation of the physical law and the 
value of said law if it was perfectly satisfied which 
is $0$ in most cases.  Again, one must be careful in 
cases like the linear and nonlinear shallow water 
equations with multiple physical laws whose violations 
differ significantly in their magnitude.  In such cases, 
compute the physical violations separately and multiply 
each term by some weight before combining them and 
computing the MSE of the combined term to 
obtain $\mathcal{L}_{phys}$.

$\mathcal{L}_{IC}$ allows the PINO to learn how the 
input initial condition given to the network is the 
value of the output at that point at $t=0$.  Moreover, by 
minimizing $\mathcal{L}_{IC}$, $\mathcal{L}_{phys}$ 
more easily converges to the correct solution for 
cases where the physical law is a PDE.  We 
defined $\mathcal{L}_{IC}$ as the relative MSE 
between the PINO prediction at $t=0$ and the 
initial condition fed into the network.  There were 
no cases where the ICs differed significantly in 
magnitude for the PDEs since for the linear and 
nonlinear shallow water equations, the velocity 
fields were taken to be zero at $t=0$ and were 
not fed into the initial data.

To combine these terms into the total 
loss $\mathcal{L}_{tot}$, we perform a weighted 
sum defined as

\begin{align}
    \label{eq:loss}
    \mathcal{L}_{total} &= w_{data} \mathcal{L}_{data} + w_{phys} \mathcal{L}_{phys} + w_{IC} \mathcal{L}_{IC}\,,
\end{align}

\noindent where $w_{data}$ is the data weight, $w_{phys}$ 
is the physics weight, and $w_{IC}$ is the IC weight.  
We varied these values between different cases and 
even during the training.  Typically, we would set 
$w_{data}$ to be $5$ or $10$, $w_{phys}$ to be $1$ 
or $2$, and $w_{IC}$ to be $5$ or $10$.

{
We emphasize that we select the weights in the loss function through a process of trial and error
rather than a sophisticated hyperparameter optimization process.
This is typically the case with PINNs as well.  If we take some limits,
we can understand why this is the case.  If $w_{phys}$ is very high,
the PINO will try to minimize $\mathcal{L}_{phys}$ at the cost of the other
parameters. One solution that satisfies the most PDEs is if the output field
is zero for all space and time coordinates.  Therefore, we add the initial condition
loss $\mathcal{L}_{IC}$ to ensure that $\mathcal{L}_{phys}$ evolves the correct
initial data.  However, if $\mathcal{L}_{IC}$ is too large, the output field
is correct at $t=0$, but it does not evolve with time.  Thus, we used a process
of trial and error to determine the correct weights.
}

{
For the training itself, we used the \texttt{PyTorch} framework \cite{paszke2019pytorch}.  We employed an Adam optimizer with $\beta_1=0.9$, $\beta_2=0.999$ with an initial learning rate of 0.001.  To fine tune the latter training steps, we employed a multistep scheduler to reduce the learning rate at several intervals throughout the training with a gamma value of 0.5.  The specific number of epochs and the epoch decay milestones vary between the different test cases.  A common setup that was used from several of the 2D cases was to train for 150 epochs and put the scheduler milestones at [25, 50, 75, 100, 125, 150] epochs.  To further improve the performance, we would use the checkpoints after 150 epochs of training and retrain again after for this same duration.  In between restarting from these checkpoints, we would sometimes modify the weights of the various training losses if we observed some losses were lagging behind the others.
}

{
\subsection{Model architecture}
As noted in \ref{sec:PINO}, the model uses the FNO architecture
that is shown in Figure~\ref{fig:architecture}.  We used a Gaussian 
error linear units (GELUs) for our nonlinearity for all cases~\cite{hendrycks2016GELU}.
We note that for physics informed deep learning, we require the nonlinear activation
function to have a non-zero second derivative, ruling some commonly used 
activation functions such as the rectified linear unit (ReLU)~\cite{raissi2019physics}.

The FNO architecture can be described by the widths, modes,
and number of layers. The widths and modes are given as arrays 
representing the width and modes for each entry.  The size of the 
array gives the number of layers and each entry corresponds to
a different layer.  We used 4 layers in all cases.
For the 1D cases, we used the $width=[16, 24, 24, 32]$ and
$modes=[15, 12, 9, 9]$.  For the 2D cases, we used
$width=[64, 64, 64, 64]$ and $modes=[8, 8, 8, 8]$.

In addition, the network ends with a fully connected layer.
We selected a width of 128 for the fully connected layer for all cases.
}

\subsection{Performance quantification}
To quantify the performance, we ran PINO on the test 
dataset, then calculated the MSE of their predictions 
with the test data and the MSE of the physics loss 
that accompanied violations of physical laws.  The test 
dataset is composed of approximately $10\%$ of the 
simulations produced from the random initial data that 
was separated from the rest of the data to ensure 
that the PINO did not train on it.

\section{Test Problems}
\label{sec:Tests}

We use PINOs to learn nine different PDEs. We consider 
the wave equation in 1D and 2D to demonstrate that our 
methods produce accurate and reliable results. We also 
present results for the 1D Burgers equation, which was 
used in~\cite{PINO_li_anima} to quantify the 
performance of PINOs to learn PDEs. We then put 
our methods at work 
to solve a variety of PDEs with different levels 
of complexity.

{
Specifically, many past works including
\cite{Li2020GraphNO,Li2020MGNO,Li2021FourierNO,kovachki2021neuraloperator,PINO_li_anima}
investigating using nonlinear neural operators study only 3 cases, 
the 1D Burgers equation, 2D Darcy Flow, and the 2D Navier Stokes equations.
While these cases provide a good way to compare neural operators to each other, 
they limited the applicability of neural operators to other problems.
Moreover, the aforementioned cases fail to include various physical and numerical
phenomena such as shocks, coupled PDEs, and non-constant coefficients.
Therefore, we chose a wide array of linear and nonlinear PDEs
to evaluate the PINO models and isolate the places where they may have difficulty.
}

\subsection{Wave Equation 1D}
\label{sec:Wave1D_description}

Our first test was the wave equation in 1D with 
periodic boundary conditions.  This is a 
computationally simple PDE that is second order 
in time and models a variety of different physics 
phenomena.  The equation for the evolved field $u(x,t)$ takes the form

\begin{align}
    u_{tt}  + c^2 u_{xx}&=0,  \label{eq:wave1d}\\
    u\left( x, 0 \right) &= u_0\left(x\right)\,, \nonumber \\
    x\in \left[ 0,1 \right),&\ t\in \left[0, 1 \right]\,, \nonumber
\end{align}

\noindent where $c=1$ is the speed of the wave.
\subsection{Wave Equation 2D}
\label{sec:Wave2D_description}
We extend the wave equation in 2D with periodic 
boundary conditions to explore the requirements 
for adding the additional spatial dimension. 
The equation for the evolved field $u(x,y,t)$ is given by
\begin{align}
 \label{eq:wave2d}
    &u_{tt}  + c^2 \left( u_{xx} + u_{yy}\ \right)=0 \\
    &u\left( x,y, 0 \right) = u_0\left(x,y\right)\,, \nonumber \\
    \nonumber 
    &x,y\in \left[ 0,1 \right),\ t\in \left[0, 1 \right] \,,
\end{align}
\noindent where as before the speed of the wave is set to $c=1$.

{
\subsection{Wave Equation 2D Non-Constant Coefficients}
\label{sec:Wave2D_nonconst_description}
By adding a spatially variable wave speed, we study the performance
of PINOs in problems with non-constant coefficients.
This variable wave speed $c(x,y)$ was incorporated into the PINO
by treating it as another randomly generated input field.  
To produce smoother training data, we set the spatial scale
$l=0.5$ parameter for the wave speed input field $c(x,y)$, 
though we still used $l=0.1$ for the initial data.
The equation for the evolved field $u(x,y,t)$ is now given by
\begin{align}
 \label{eq:wave2d_nonconst}
    &u_{tt}  + c(x,y)^2 \left( u_{xx} + u_{yy}\ \right)=0 \\
    &u\left( x,y, 0 \right) = u_0\left(x,y\right)\,, \nonumber \\
    \nonumber 
    &x,y\in \left[ 0,1 \right),\ t\in \left[0, 1 \right] \,.
\end{align}
}

\subsection{Burgers Equation 1D}
\label{sec:Burgers1D_description}
The 1D Burgers equation with periodic boundary 
conditions serves as a nonlinear test case with for 
a variety of numerical methods.  This allowed us to 
verify that our PINOs can learn and reconstruct 
nonlinear phenomena.  The equation for the field $u(x,y,t)$ is given in 
conservative form by

\begin{align}
\label{eq:burgers1d} 
u_{t}+\partial_{x}\left(u^{2} / 2\right) &=\nu u_{xx}, \\ 
\nonumber
u(x, 0) &=u_{0}(x)\,, \\ \nonumber
x \in[0,1),& \ t \in[0,1]\,, 
\end{align}
\noindent where the viscosity $\nu=0.01$.

\subsection{Burgers Equation 2D Scalar}
\label{sec:Burgers2D_description}

To verify our model can handle nonlinear phenomena in 2D, 
we extend the Burgers equation into 2D by assuming 
the field $u(x,y,t)$ is a scalar.  The equations take the form

\begin{align}
\label{eq:burgers2d} 
u_{t}+\partial_{x}\left(u^{2} / 2\right) + \partial_{y}\left(u^{2} / 2\right) &=\nu \left(u_{xx} +u_{yy}\right)\,, \\ \nonumber
u(x, y, 0) &=u_{0}(x, y)\,, \\
x,y \in[0,1),& \ t \in[0,1]\,, \nonumber\
\end{align}
\noindent where the viscosity $\nu=0.01$.

\subsection{Burgers Equation 2D Inviscid}
\label{sec:Burgers2DInvisc_description}
We also looked at cases involving the inviscid 
Burgers equation in 2D in which we set the viscosity 
$\nu=0$.  This setup is known to produce shocks 
that can result in numerical instabilities if not 
handled correctly.  We used a finite volume method (FVM) 
to generate this data to ensure stability in the presence 
of shocks.  In turn, this allowed us to investigate 
the network's performance when processing shocks. 
The equations are given by

\begin{align}
\label{eq:burgers2d_inviscid} 
u_{t}+\partial_{x}\left(u^{2} / 2\right) + \partial_{y}\left(u^{2} / 2\right) &=0\,,\\
\nonumber
x,y \in[0,1),& \ t \in[0,1]\,,  \\
u(x, y, 0) &=u_{0}(x, y). \nonumber
\end{align}

\noindent We observed that the presence of the 
shock prevented our Fourier derivative method from 
producing accurate residuals when we included the 
full equation in our physics loss term.  Instead, 
for our physics loss, we used the conserved quantity,

\begin{align}
\label{eq:burgers2d_inviscid_cons_law}
\int_{\Omega}u(x,y,t) \ dx\ dy = \int_{\Omega}u(x,y,0) \ dx\ dy = C\,,
\end{align}

\noindent where $C$ is a constant and $\Omega$ is 
the domain.  In other words, we ensured that the 
total $u$ at every time instance is equal to the total 
$u$ at $t=0$.

\subsection{Burgers Equation 2D Vector}
\label{sec:Burgers2DVec_description}
We then looked at a vectorized form of the 2D 
Burgers equation with periodic boundary conditions. 
This allowed us to test how well the model handles 
the coupled fields $u(x,y,t)$ and $v(x,y,t)$ that parameterize the 
system. The equations take the form

\begin{align}
\label{eq:burgers2d_vec_I} 
u_{t}+u u_{x} + v u_{y} = &\nu \left(u_{xx} +u_{yy} \right)\,, \\
\label{eq:burgers2d_vec_II} 
v_{t}+u v_{x} + vv_{y} = &\nu \left(v_{xx} +v_{yy} \right)\,, \\
u(x, y, 0) =u_{0}(x, y),&\ v(x, y, 0) = v_{0}(x, y), \nonumber\\  
x,y \in[0,1),& \ t \in[0,1] \nonumber
\end{align}

\noindent where the viscosity $\nu=0.01$.  We note that 
this system of equations does not have a conservative 
form as there is not a continuity equation in this 
system.

{
Although coupled equations might seem like a trivial case,
there are actually a number of complexities that arise when
changing the number of fields.  First, having multiple inputs
and outputs results in an expanded parameter space for the PINO.
Thus, one would expect the network to require a larger volume
of training data to produce accurate results.  Moreover,
PINOs must solve multiple equations simultaneously in the coupled case.
In turn, the models must not only solve for single fields,
but also compute the contribution of those fields on the other fields.
Therefore, it is important to understand how PINOs can resolve coupled
fields in this relatively simple 2D vectorized Burgers equation.
}

\subsection{Linear Shallow Water Equations 2D}
\label{sec:ShallowWaterLinear_description}
To examine the properties of PINOs with 3 coupled 
equations, we examined the ability of the networks 
to reproduce the linear shallow water equations with 
periodic boundary conditions.  We assumed that the 
height of the perturbed surface $h(x,y,t)$ is 
initially perturbed, but the initial velocity fields
$u(x,y,t)$ and $v(x,y,t)$ are 
initially zero. These equations can be expressed as

\begin{align}
\label{eq:swe_lin_I}
&\frac{\partial h}{\partial t}+H\left(\frac{\partial u}{\partial x}+\frac{\partial v}{\partial y}\right)=0\,, \\
\label{eq:swe_lin_II}
&\frac{\partial u}{\partial t}-f v=-g \frac{\partial h}{\partial x}\,, \\
\label{eq:swe_lin_III}
&\frac{\partial v}{\partial t}+f u=-g \frac{\partial h}{\partial y}\,,
\end{align}
\begin{align}
&\textrm{with} \quad h(x,y,0) = h_{0}(x,y),\ u(x,y,0)=0,\ v(x,y,0)=0,\ \nonumber \\ 
&x,y \in[0,1), \ t \in[0,1], \nonumber
\end{align}

\noindent where the gravitational constant $g=1$, the mean fluid height $H=100$, and we considered two cases for the 
Coriolis coefficient $f=\{0,1\}$. 

{
The challenge in this case is that we have 3 coupled fields
with one of them, the perturbed surface height $h$ having a very different physical 
meaning than the others.  Moreover, $h$ is typically of much larger magnitude than
either of the velocity fields.  By simplifying to a linear problem,
we assess the ability of PINOs to reproduce the results of magnitude
varying fields without complicated nonlinear terms.  We note that coupled
equations with terms of varying magnitude are known to be difficult for 
traditional PINNs as one needs to carefully weight and normalize the
equations.
}

\subsection{Nonlinear Shallow Water Equations 2D}
\label{sec:ShallowWaterNonlinear_description}
Finally, we examined the network performance on the 
nonlinear shallow water equations.  
We assumed a similar setup as in the linear case where we assumed the total fluid column height $\eta(x,y,t)$, was given by a mean value of $1$ plus some initial perturbation.  We again assumed the initial velocity fields $u(x,y,t)$ and $v(x,y,t)$ were zero.  These equations are given by

\begin{align}
\label{eq:swe_nonlin_I}
\frac{\partial(\eta)}{\partial t}+\frac{\partial(\eta u)}{\partial x}+\frac{\partial(\eta v)}{\partial y}&=0\,,  \\
\label{eq:swe_nonlin_II}
\frac{\partial(\eta u)}{\partial t}+\frac{\partial}{\partial x}\left(\eta u^{2}+\frac{1}{2} g \eta^{2}\right)+\frac{\partial(\eta u v)}{\partial y}&=\nu\left(u_{xx} + u_{yy}\right)\,, \\
\label{eq:swe_nonlin_III}
\frac{\partial(\eta v)}{\partial t}+\frac{\partial(\eta u v)}{\partial x}+\frac{\partial}{\partial y}\left(\eta v^{2}+\frac{1}{2} g \eta^{2}\right)&=\nu\left(v_{xx} + v_{yy}\right)\,,
\end{align}
\begin{align}
&\textrm{with} \quad \eta(x,y,0) = \eta_{0}(x,y),\ u(x,y,0)=0,\ v(x,y,0)=0,\ \nonumber \\ 
&x,y \in[0,1), \ t \in[0,1], \nonumber
\end{align}

\noindent where the gravitational coefficient $g=1$ and the viscosity coefficient $\nu=0.002$ to prevent the formation of shocks.

{
This case combines the difficulty of having 3 coupled fields with complicated
nonlinear governing equations.  This case provides a very useful and physically
interesting benchmark for PINOs as these equations model tsunamis.  Moreover,
these equations take a similar form to the equations of compressible flow,
but without an additional equation for energy that is dependent on the 
equation of state.
}

All these different PDEs serve the purpose 
of establishing 
the accuracy and reliability of our PINOs, and then 
explore their application for more interesting scenarios 
for the 2D Burgers equation, and 2D linear and nonlinear 
shallow waters equations, which involve several coupled 
equations. 

\begin{figure}[htb]
    \centering
    \includegraphics[width=\linewidth]{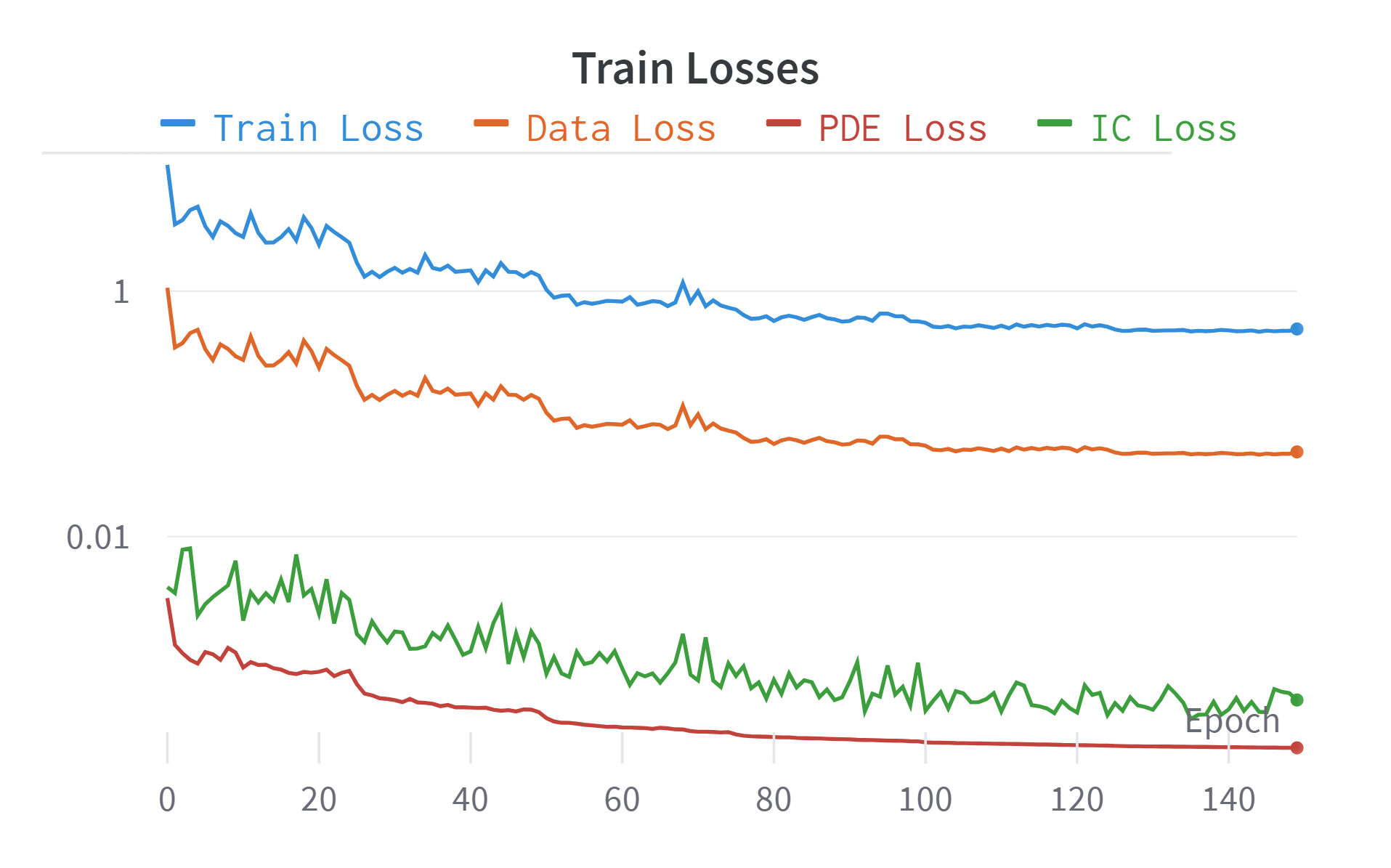}
    \caption{The training loss curves for the first 150 epochs of training for the nonlinear shallow water equations.  Here we compare the different losses from the data, PDE, and ICs as well as their weighted total training loss. }
    \label{fig:training_curves}
\end{figure}

\begin{table*}[htb]
    \centering
    \caption{\textbf{Summary of PINO results} The first column describes the modeled equation.  The second and third columns describe the spatial and temporal resolution, respectively.  The fifth and sixth columns display the number of training and testing samples used.  The final column provides the relative mean squared error (MSE) of our physics informed neural operators on the test set.}
    \begin{tabular}{| p{6.5cm} | p{1.55cm} | p{1cm} | p{1.5cm} | p{1.5cm} | p{1.5cm} |}
        \toprule
        Model & Spatial Resolution & Time Steps & Training Samples & Testing Samples & Relative MSE \\
        \midrule
        Wave Equation 1D & 128   & 101   & 900   & 100   & 1.22E-03 \\
        Wave Equation 2D & 128 $\times$ 128 & 101   & 45    & 25     & 6.60E-03 \\
        Wave Equation 2D Non-Constant Coefficients & 128 $\times$ 128 & 101 & 175 & 25 & 4.86E-02 \\
        Burgers Equation 1D & 128   & 101   & 90    & 10    & 6.94E-03 \\
        Burgers Equation 2D Scalar & 128 $\times$ 128 & 101   & 45    & 25     & 3.56E-03 \\
        Burgers Equation 2D Scalar FNO & 128 $\times$ 128 & 101   & 45    & 25     & 6.29E-03 \\
        Burgers Equation 2D Inviscid & 128 $\times$ 128 & 101   & 90    & 10    & 3.56E-02 \\
        Burgers Equation 2D Vector & 128 $\times$ 128 & 101   & 475   & 25    & 8.49E-03 \\
        Linear Shallow Water Equations 2D f=0 & 128 $\times$ 128 & 101   & 5     & 25     & 3.61E-02 \\
        Linear Shallow Water Equations 2D f=1 & 128 $\times$ 128 & 101   & 45    & 25     & 9.19E-03 \\
        Nonlinear Shallow Water Equations 2D & 128 $\times$ 128 & 101   & 45    & 25     & 1.50E-02 \\
        \bottomrule
        \end{tabular}%
    \label{tab:mse}%
\end{table*}%

\section{Results}
\label{sec:results}

\begin{figure*}[htb]
\includegraphics[width=\textwidth]{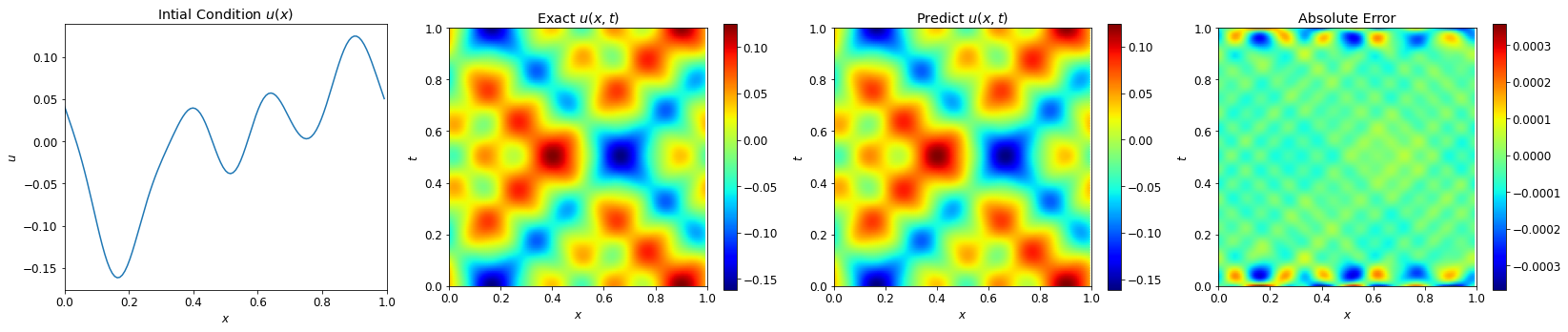}
\includegraphics[width=\textwidth]{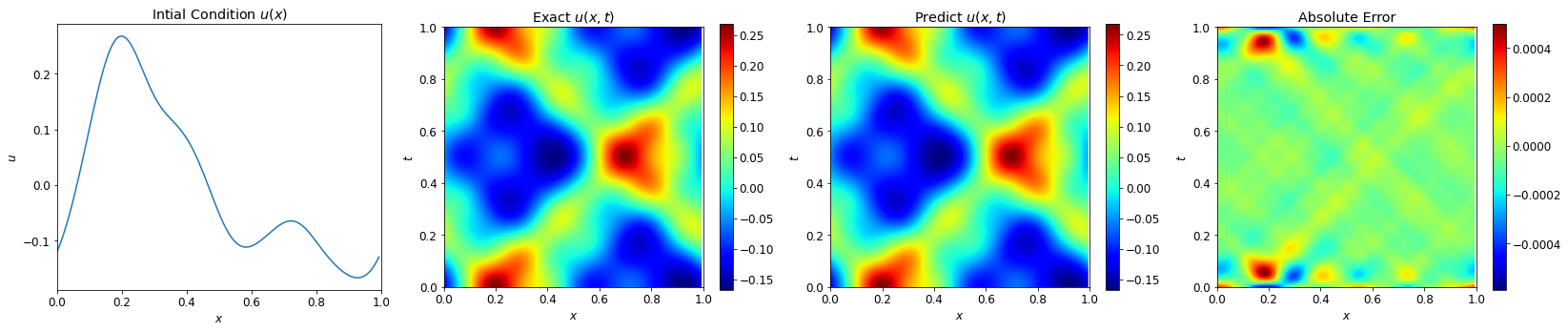}
\centering
\caption{\textbf{PINO for 1D wave equation} Test set 
initial conditions (left column) fed into 
our neural networks. The center left column 
displays the ground truth value of $u_{true}(x,t)$ 
as produced by our simulations.  The center right column 
shows the value of $u_{pred}(x,t)$ predicted by our 
PINO. The right column illustrates the error 
between the PINO predictions and the ground 
truth, $u_{pred} - u_{true}$.}
\label{fig:Wave1D}
\end{figure*}

\begin{figure*}[htb]
\includegraphics[width=\textwidth]{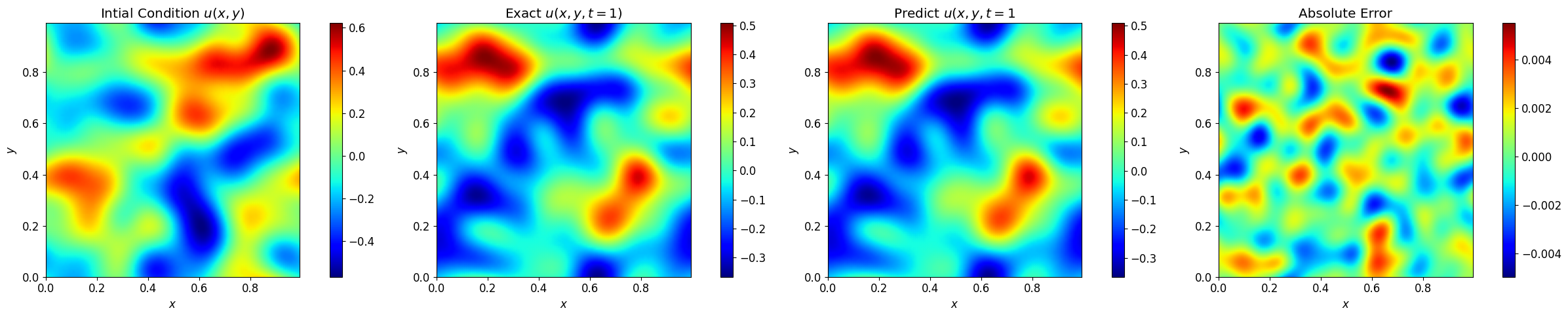}
\includegraphics[width=\textwidth]{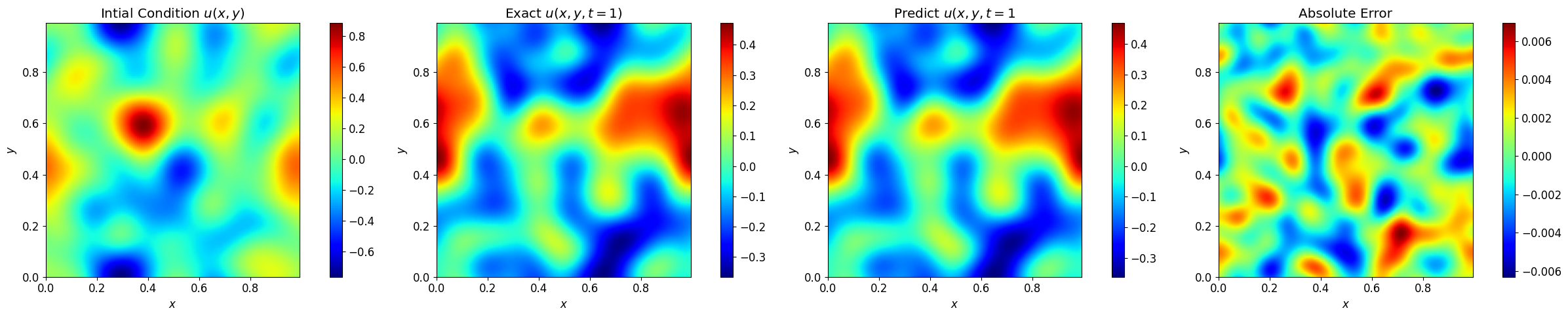}
\centering
\caption{\textbf{PINO for 2D wave equation} 
As Figure~\ref{fig:Wave1D}, but now in 2D. The left column represents the test set initial 
conditions that we feed into our PINOs. We 
evolved the systems until \(t=1\) and present 
ground truth solutions (center left) and 
PINO predictions (center right). Even 
after evolving these simulations until the 
end of the time domain under consideration, 
our PINOs predict with excellent accuracy 
the evolution of the system, as shown 
in the right column which shown at $t=1$, $u_{pred} - u_{true}$.}
\label{fig:Wave2D}
\end{figure*}

We now quantify the ability of our PINOs to learn the 
physics described by the PDEs described above. In 
Figures~\ref{fig:Wave1D}-\ref{fig:swe_nonlin} we present 
qualitative and quantitative results that illustrate 
the performance of our PINOs. We use two types of 
quantitative metrics, namely, absolute error (shown in 
each figure) and 
mean squared error (summarized in Table \ref{tab:mse}) 
between PINO predictions and ground truth simulations. 
While the figures below provide snapshots of the performance 
of our PINOs at a given time, \(t\), we also provide 
interactive visualizations of these results 
at this 
\href{https://shawnrosofsky.github.io/PINO_Applications/}{website}.
We also provide a tutorial to use our PINOs and 
reproduce our results in the 
\href{https://www.dlhub.org}{Data and Learning Hub for Science}~\cite{dlhub,blaiszik_foster_2019}.

{
Before discussing specific results, we also present the training curves of the nonlinear shallow water equations over its first 150 epochs in Figure~\ref{fig:training_curves}.  Once our PINOs are fully trained, they are computationally efficient. Averaging over 25 test sets, our PINOs produce full simulations within 0.165 seconds for the most computationally intensive case, the nonlinear shallow water equations.
}

\subsection{Wave Equation 1D Results}
\label{sec:Wave1D_results}
Figure~\ref{fig:Wave1D} shows that our PINO for 
the 1D wave equation learns and describes the 
physics of this PDE with remarkable accuracy. 
These results serve the purpose 
of validating our methods with a simple PDE.
We found that the MSE in this case was 
$1.22\times 10^{-3}$ on the test dataset. We note 
that for this case, we experimented with using a 
very high resolution of $4096$ grid points to 
generate the training data, but downsampled by 
a factor of $32$ to a final resolution of $128$ 
grid points that were fed into the network. 
This simulates having available less data 
to reproduce a result than was required to compute it.  

\subsection{Wave Equation 2D Results}
\label{sec:Wave2D_results}
Next we consider the 2D wave equation. 
Figure~\ref{fig:Wave2D} summarizes 
our results. 
As in the 1D scenario, our 
PINOs have learned the physics of the 2D 
wave equations with excellent accuracy.
Specifically, we found the MSE on the test dataset
for this case to be $6.70 \times 10^{-3}$. 

\begin{figure*}[htb]
\includegraphics[width=\textwidth]{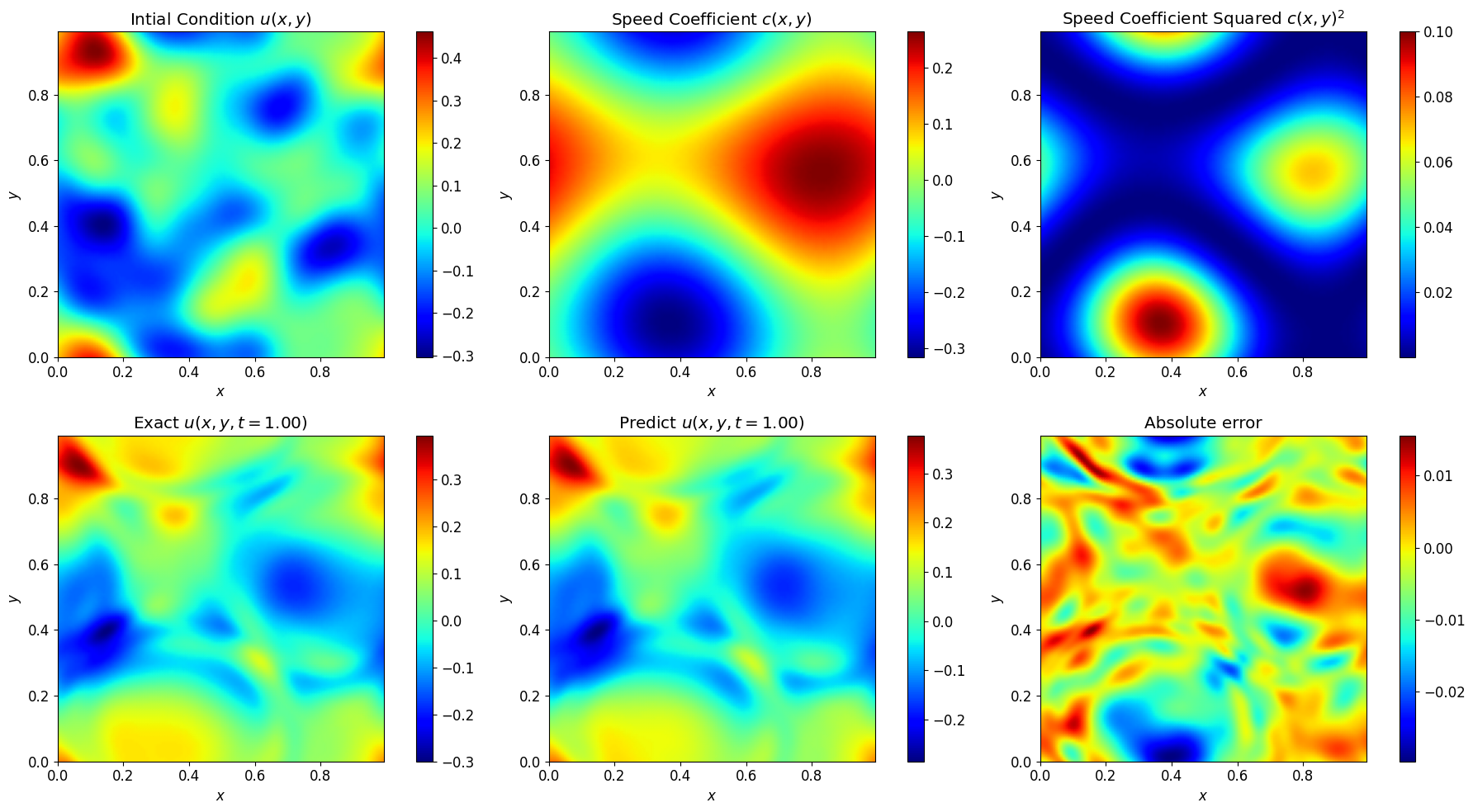}
\centering
\caption{\textbf{PINO for 2D non-constant coefficients wave equation} 
Similar to Figure~\ref{fig:Wave2D}, but here we have a non-constant wave speed. 
The top row depicts the input fields into the ANN, which are
the initial condition $u$, the wave speed coefficient $c$,
and the wave speed coefficient squared $c^2$ 
from left to right respectively.
As in the constant coefficient case, we
evolved the systems until \(t=1\).
On the bottom row, we present the values at $t=1$ 
 for the ground truth solutions (left), 
the PINO predictions (center), the error $u_{pred}-u_{true}$.}
\label{fig:Wave2D_nonconst}
\end{figure*}

{
\subsection{Wave Equation 2D Non-Constant Coefficients Results}
\label{sec:Wave2D_nonconst_results}

Next we consider the variation of the 2D wave equation
with non-constant coefficients.
Figure~\ref{fig:Wave2D_nonconst} illustrates the results
for this problem.
As this scenario is considerably more difficult than the
uniform coefficient wave equation, the 
PINO performance is slightly worse,
with an MSE of on the test dataset of $0.0486$.

\begin{figure*}[htb]
\includegraphics[width=\textwidth]{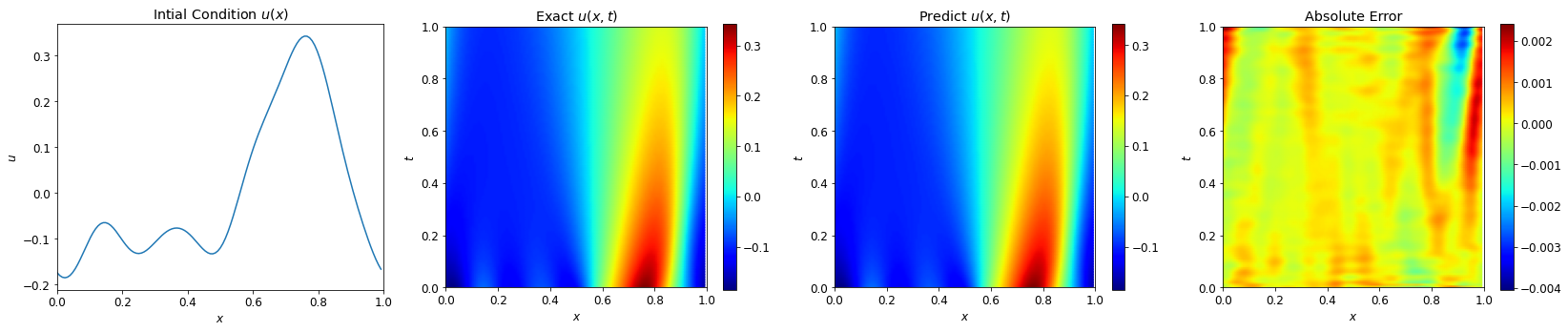}
\includegraphics[width=\textwidth]{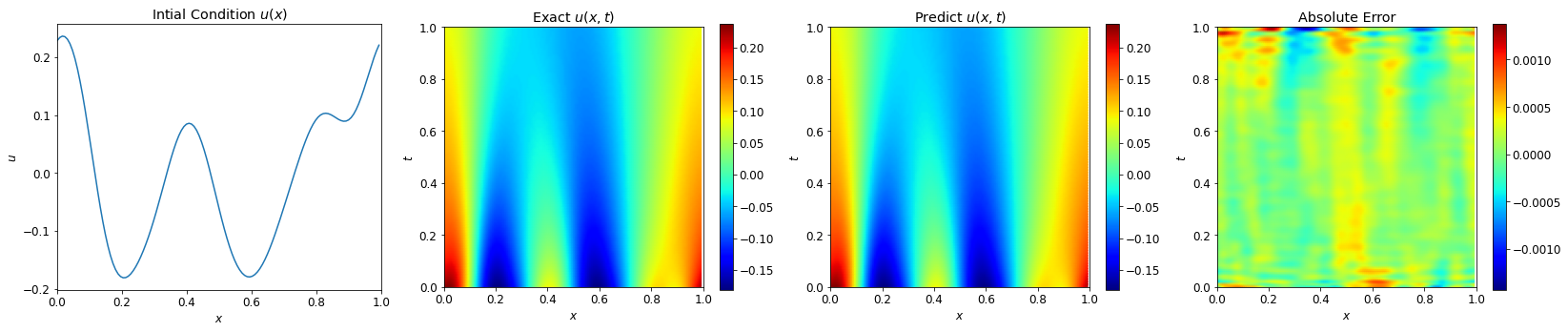}
\centering
\caption{\textbf{PINO for 1D Burgers equation} The left 
column shows the test set initial conditions fed into 
our PINOs. The center left column shows the ground 
truth value, $u_{true}(x,t)$, produced by our simulations. 
The center right column presents the predicted 
values $u_{pred}(x,t)$ by our PINOs. The right 
column shows the discrepancy between PINO 
and ground truth predictions, $u_{pred} - u_{true}$.}
\label{fig:Burgers1D}
\end{figure*}
}
\subsection{Burgers Equation 1D Results}
\label{sec:Burgers1D_results}

We now turn our attention to the Burgers equation, and 
begin this analysis with a simple and illustrative case, 
namely the 1D Burgers equation, shown in Figure~\ref{fig:Burgers1D}. These results show that 
our PINOs have accurately learned the physics described by 
this PDE with an excellent level of accuracy.
Quantitatively, the MSE on the test dataset was $6.94 \times 10^{-3}$.
These results furnish evidence that our methods can 
reproduce results published 
elsewhere in the literature~\cite{PINO_li_anima}.

We now turn our attention to several 2D Burgers 
equations that, to the best of our knowledge, have not 
been explored previously in the literature.

\begin{figure*}[htb]
\includegraphics[width=\textwidth]{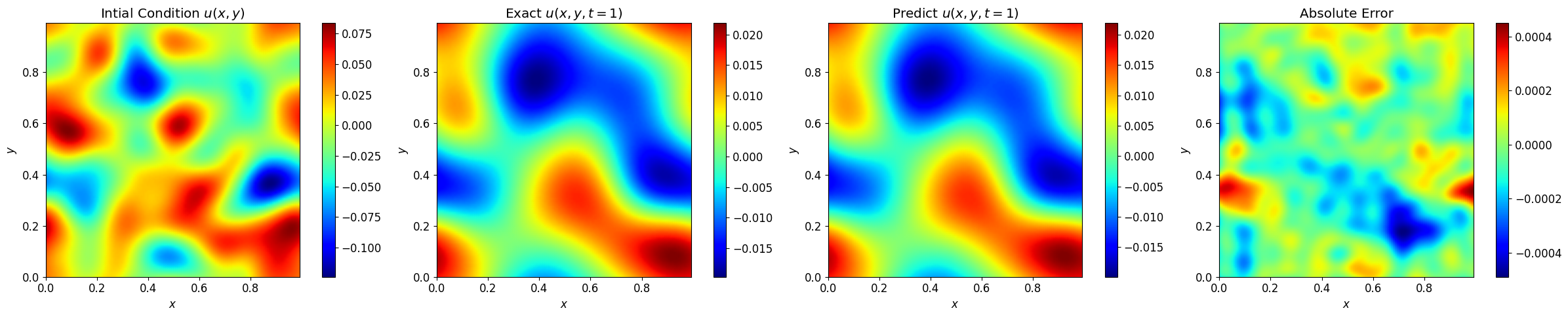}
\includegraphics[width=\textwidth]{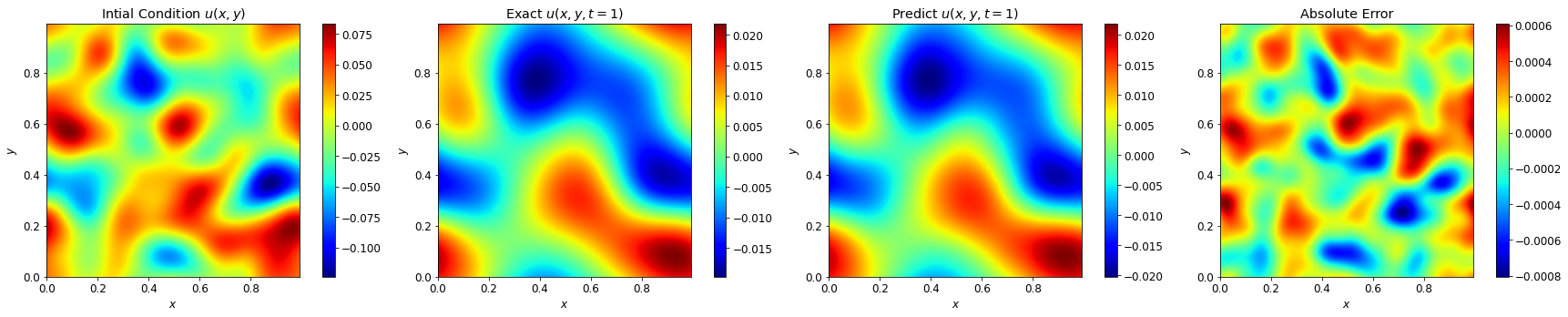}
\centering
\caption{\textbf{PINO and FNO for 2D scalar Burgers equation} 
Displays the results of the PINO and FNO models for
this equation.
The left column shows a sample of test set initial conditions 
fed into our models. The center left column displays 
the ground truth value, $u_{true}(x,y)$, of our 
simulations once they have evolved up to \(t=1\). 
The center right column shows model predictions for this 
PDE at \(t=1\). We have selected this time to quantify 
the accuracy of our model once the system has evolved 
sufficient time to accumulate errors. The right 
column illustrates the discrepancy between model 
predictions and the ground truth, $u_{pred} - u_{true}$, 
at \(t=1\).  The top row shows the results for the PINO and the bottom 
row shows the results for the FNO.}
\label{fig:Burgers2D}
\end{figure*}

\subsection{Burgers Equation 2D Scalar Results}
\label{sec:Burgers2D_results}

This PDE is given by 
Equation~\eqref{eq:burgers2d}. As shown 
in Figure~\ref{fig:Burgers2D} our PINOs can learn 
and describe the physics of this PDE accurately. 
Note that we present results for this PDE once the 
system has been evolved throughout the time domain 
under consideration, i.e., \(t\in[0,1]\). By presenting 
results at \(t=1\) we gain a good understanding of the 
actual performance of our PINOs once they have evolved 
in time and accumulated numerical errors that may 
depart from ground truth values. 
{
For this case, we compared the results to a FNO
which was trained with the same architecture,
but without a physics informed loss component.
In Figure~\ref{fig:Burgers2D} we present the results
of the PINO and the FNO for the same initial condition.
We found that the MSE for the full test dataset
was $3.56 \times 10^{-3}$. 
for the PINO and was 
$6.29 \times 10^{-3}$ for the FNO.
}

\begin{figure*}[htb]
\includegraphics[width=\textwidth]{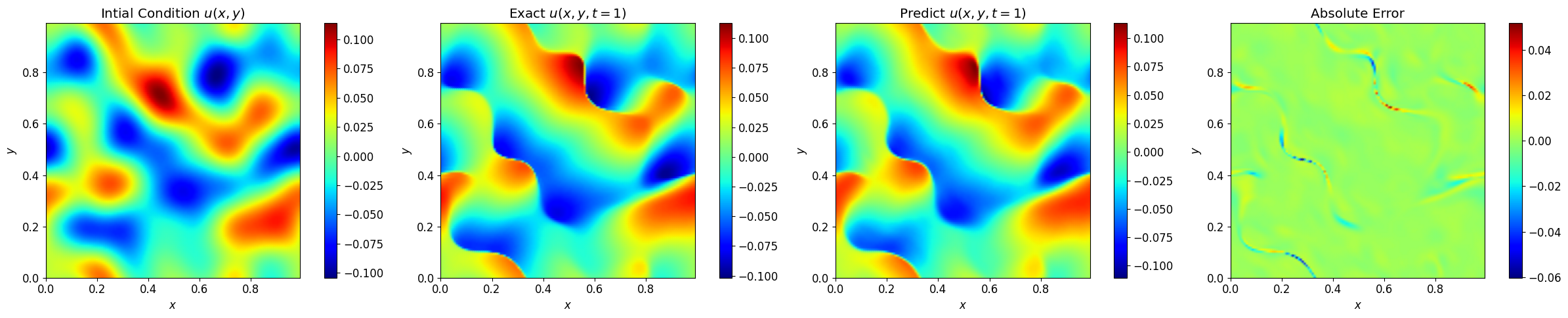}
\includegraphics[width=\textwidth]{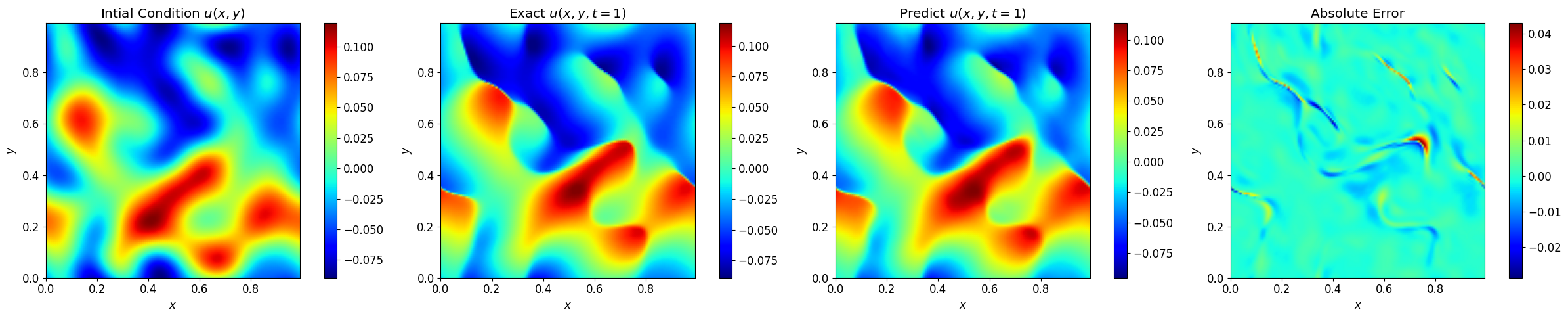}
\centering
\caption{\textbf{PINO for 2D inviscid Burgers equation} As Figure~\ref{eq:burgers2d} but now in the inviscid case 
given by Equation~\ref{eq:burgers2d_inviscid}.}
\label{fig:Burgers2DInvisc}
\end{figure*}

\subsection{Burgers Equation 2D Inviscid Results}
\label{sec:Burgers2DInvisc_results}
This PDE is given 
by Equation~\ref{eq:burgers2d_inviscid}. As we mentioned before, 
shocks are a common occurrence for this PDEs, and may 
lead to numerical instabilities if we do not use 
shock capturing schemes. We have quantified the 
ability of our PINOs to handle shocks. 
We found the PINO to have an MSE of $0.0356$ on the test dataset.
In Figure~\ref{fig:Burgers2DInvisc} we notice 
that our PINOs are able to describe the physics of this 
PDE with excellent accuracy. However, we observe  a 
significant discrepancy between ground truth and AI 
predictions right at the regions where shocks occur. 
These findings indicate that further work is needed 
to better handle PDEs that involve shocks. This is 
a specific area of work that we will pursue in 
the future. 

{ 
Specifically, such work would look at improvements to the way the
model and loss function handle discontinuities in the data.
For example, Fourier transforms, which are employed in the neural network
and to represent derivatives in the loss function, are known
to be highly sensitive to discontinuities. Thus, 
potential improvements may encompass the use of 
Galerkin neural networks~\cite{ainsworth2021galerkin} to better handle 
discontinuities, and loss functions that 
incorporate particle number conservation. These 
improvements should be considered in future work.
}

\subsection{Burgers Equation 2D Vector Results}
\label{sec:Burgers2DVec_results}
\begin{figure*}[htb]
\includegraphics[width=\textwidth]{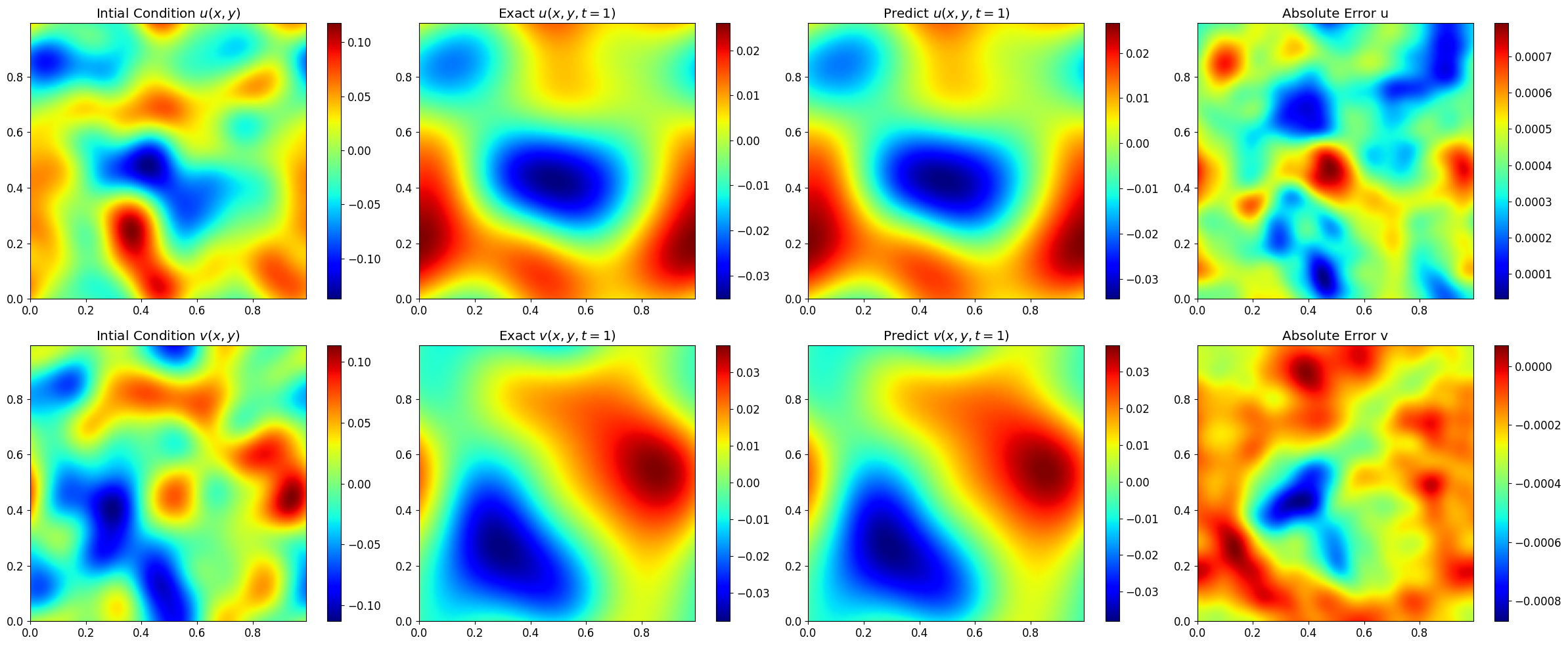}
\centering
\caption{\textbf{PINO for 2D vector Burgers equation} 
The top and bottom rows show the \(u\) and \(v\) fields, 
respectively. The left column shows test set 
initial conditions for \((u, v)\). The center left 
column shows ground 
truth values for the \((u, v)\) fields once this PDE 
has been evolved up to \(t=1\). The center right column 
shows PINO results for the \((u, v)\) fields at 
\(t=1\). The right column show the discrepancy between 
PINO and ground truth values at \(t=1\).}
\label{fig:Burgers2DVec}
\end{figure*}

This is the most 
complex PDE of the Burgers family we 
consider in this study, see 
Equations~\eqref{eq:burgers2d_vec_I} 
and~\eqref{eq:burgers2d_vec_II}. The novel feature of 
this PDE is that we now need to treat two different fields 
\((u,v)\). 
In Figure~\ref{fig:Burgers2DVec} we present
a sample result from the test dataset,
which demonstrate that our PINOs 
have learned the physics of this PDE 
and describe it with remarkable precision even after we 
have evolved this system until the end of the time 
domain under consideration.
Quantitatively, we achieved an MSE of $8.49 \times 10^{-3}$ on the test dataset.

These results complete our analysis for a variety of 
PDEs that involve the Burgers equation, and indicate that for various levels of complexity and initial conditions
our PINOs are capable of learning and describing 
the physics of these PDEs with excellent accuracy. 
We have also realized that we need to further develop 
these methods for PDEs that involve shocks. We are 
keenly interested in this particular case, 
and will explore in future work.

\begin{figure*}[htb]
\includegraphics[width=\textwidth]{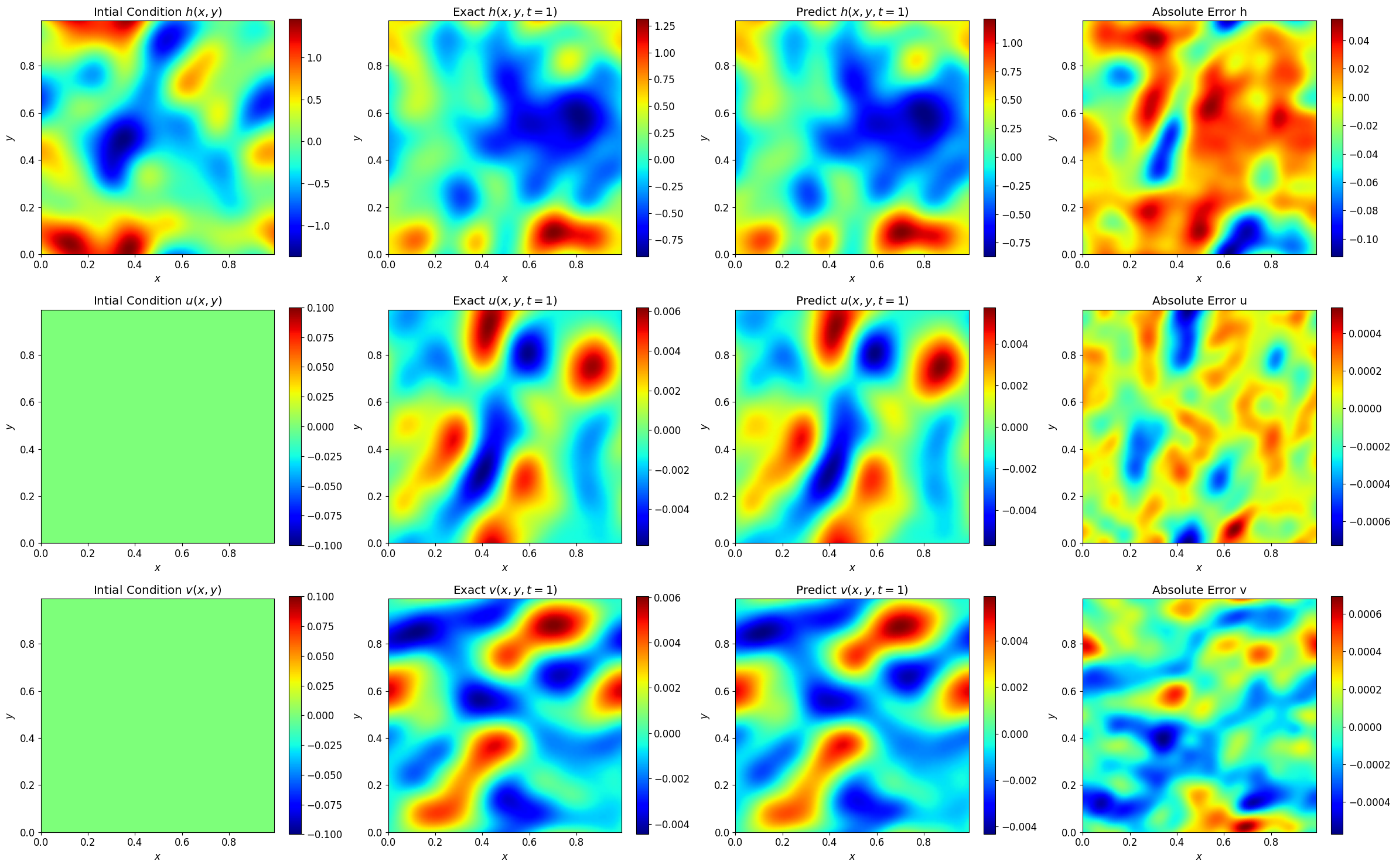}
\centering
\caption{\textbf{PINO for 2D linear shallow water equation} Assuming a system with a Coriolis coefficient $f=0$, 
we show the $h$, $u$, and $v$ fields in the top, 
middle, and bottom rows, respectively.  The left 
column shows the initial condition provided to the 
network.  The center left column displays the ground 
truth value at $t=1$ as produced by the simulation.  
The center right column shows the value of prediction at $t=1$ predicted by our PINO. The right column 
indicates the discrepancy between PINO predictions and 
the ground truth at $t=1$.}
\label{fig:swe_lin_f0_0}
\end{figure*}

\begin{figure*}[htb]
\includegraphics[width=\textwidth]{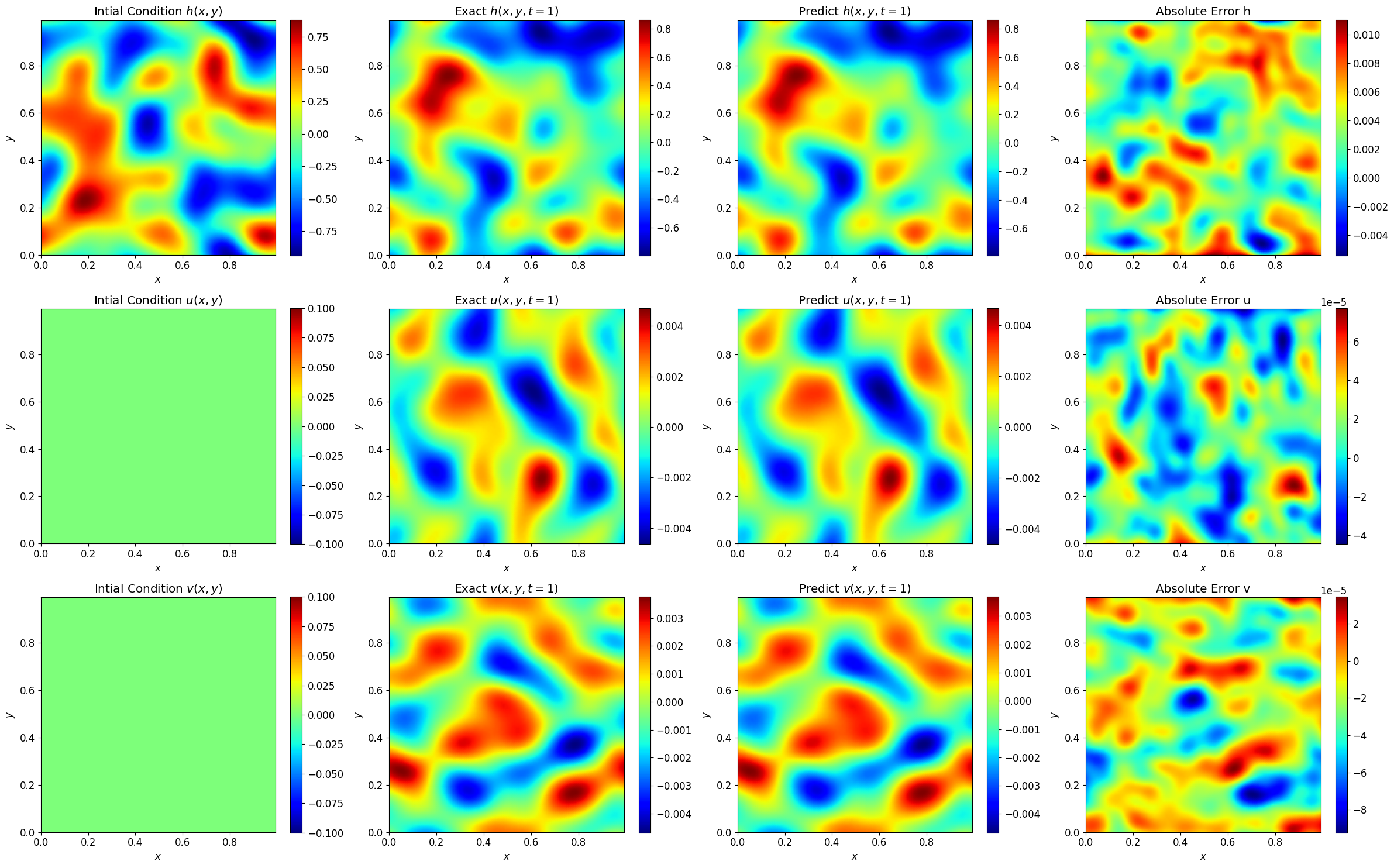}
\centering
\caption{\textbf{PINO for 2D linear shallow water equation} As Figure~\ref{fig:swe_lin_f0_0} but now with a Coriolis coefficient $f=1$.}
\label{fig:swe_lin_f0_1}
\end{figure*}

\subsection{Linear Shallow Water Equations 2D Results}
\label{sec:ShallowWaterLinear_results}

Another original result in this study is the use 
of PINOs to learn the physics of 3 coupled PDE equations. 
The first case under consideration is the 2D linear 
shallow water equation given by Equations~\eqref{eq:swe_lin_I}-\eqref{eq:swe_lin_III}. In 
this case we now consider the height of the perturbed 
surface, \(h\), and the fields \((u,v)\).  Figures~\ref{fig:swe_lin_f0_0} and~\ref{fig:swe_lin_f0_1} 
present results assuming two Coriolis coefficient 
\(f=\{0,1\}\), respectively. 
The discrepancy between PINO predictions and ground truth values 
in the figures is very small, which furnishes strong evidence for the 
adequacy of PINOs to learn the physics of this PDE
Highlights of these results include:

\begin{cititemize2}
\item \noindent \textbf{$\mathbf{f=0}$ case:} We only required 5 
training samples to produce PINOs that exhibit 
optimal performance. In this case, the MSE on the 
test dataset was $0.0361$.  This illustrates the 
ability of our PINOs to generalize from very sparse 
training data.
\item \noindent \textbf{$\mathbf{f=1}$ case:} We increased the number 
of training samples to 45, which is comparable to some 
of the other 2D cases. The MSE on the test dataset was calculated to be $9.19 \times 10^{-3}$.
\end{cititemize2}

These results provide a glimpse of the 
capabilities of PINOs to learn the physics 
of these linear, coupled PDEs. We have extensively 
tested these equations and found that they are robust 
to a broad range of initial data. 
These results provided enough motivation to explore the 
use of PINOs for a more challenging set of coupled 
PDEs, namely, the 2D nonlinear shallow water equations 
that we discuss next. 

\begin{figure*}[htb]
\includegraphics[width=\textwidth]{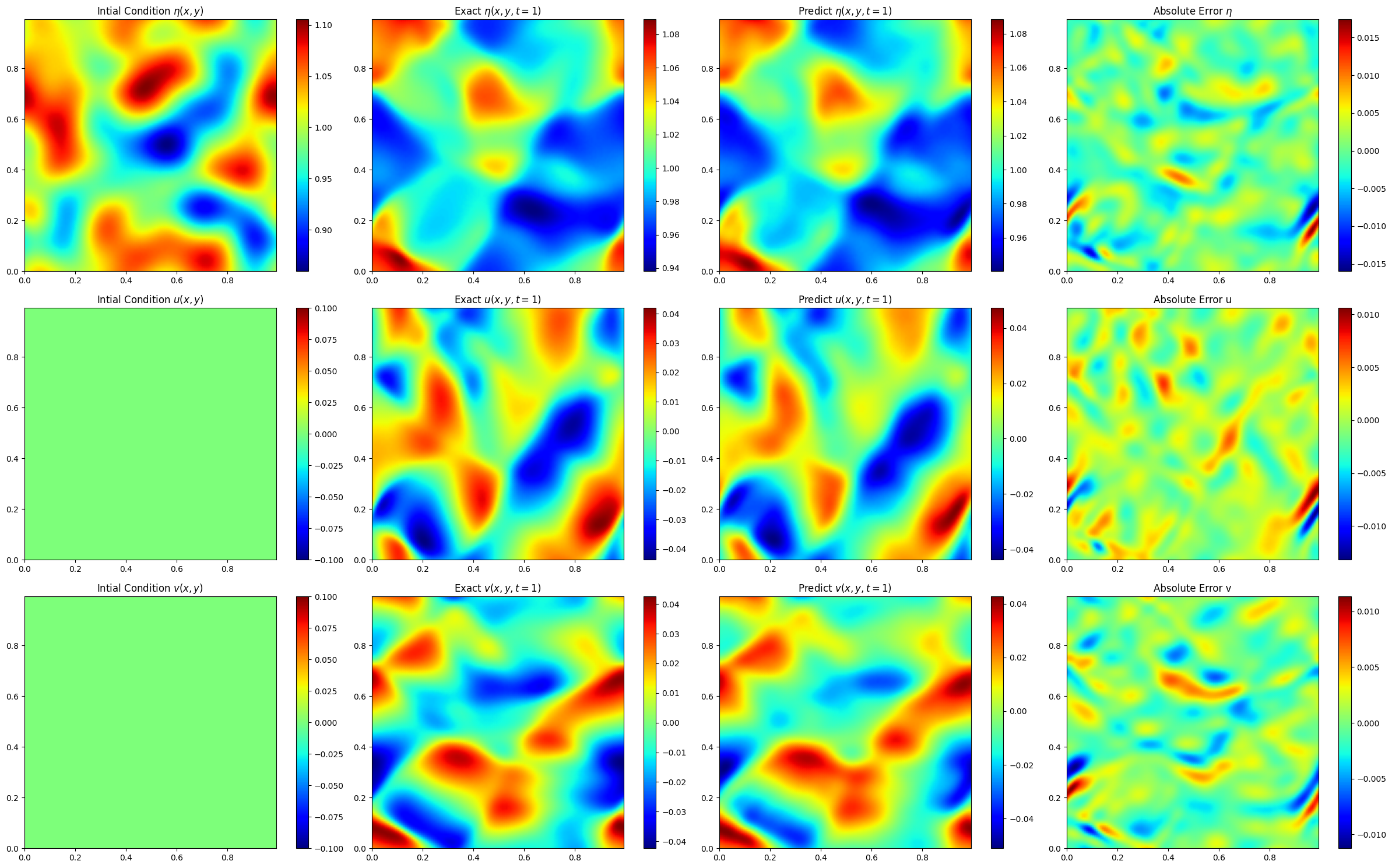}
\centering
\caption{\textbf{PINO for 2D nonlinear shallow water equation} 
The $(\eta, u, v)$ fields are shown in the 
top, middle, and bottom rows, respectively.  The left 
column presents the test set initial condition that is 
fed into our network.  The center left column displays 
the ground truth value at $t=1$ as produced by our 
simulations.  The center right column shows the value 
of prediction at $t=1$ predicted by the PINO. The right 
column illustrates the discrepancy between the 
PINO predictions and the ground truth at $t=1$.}
\label{fig:swe_nonlin}
\end{figure*}

\subsection{Nonlinear Shallow Water Equations 2D Results}
\label{sec:ShallowWaterNonlinear_results}

The final original contribution 
of this study is the use 
of PINOs to learn the physics of the 2D nonlinear shallow 
water equation, given by Equations~\eqref{eq:swe_nonlin_I}-
\eqref{eq:swe_nonlin_III}. Even though this PDE is 
significantly more complex than its linear counterpart, 
we notice in Figure~\ref{fig:swe_nonlin} that our PINOs 
learn the physics described by the fields \((\eta, u,v)\) 
with remarkable accuracy.  We found the MSE for the test dataset of this case to be 0.0150 


Finally, we provide an additional metric to quantify 
the overall performance 
of all the PDEs we have used in this study 
in Table~\ref{tab:mse}. These results indicate that our 
PINOs provide state-of-the-art results to model simple 
PDEs (1D wave equation \& 1D Burgers equations), and 
excellent performance for a variety of PDEs that are solved 
for the first time in the literature with physics informed 
neural operators.

\section{Conclusions}
\label{sec:conclusions}

We have introduced an end-to-end framework to learn 
PDEs that range from simple equations that serve the purpose 
of testing our methods (1D wave equation \& 1D Burgers equation) 
to increasingly more complex equations 
(2D scalar, 2D inviscid and 2D vector Burgers equation), and 
coupled PDEs that are solved with PINOs for the first time 
in the literature (2D linear \& nonlinear shallow waters 
equations). The methods we introduce in this study 
provide the flexibility 
to produce initial data to 
test the robustness and applicability of AI surrogates 
for a broad range of physically motivated scenarios. We 
provide scientific visualizations of our results 
through an interactive 
\href{https://shawnrosofsky.github.io/PINO_Applications/}{website}.
We also release our PINOs and scientific software through the 
\href{https://www.dlhub.org}{Data and Learning Hub for Science} 
so that AI practitioners may download, use and further 
develop our neural networks. 
In addition to this Data and Learning Hub for Science implementation,
we release the \href{https://github.com/shawnrosofsky/PINO_Applications/tree/main}{source code} 
used in this paper.

Future work may focus on the extension of these PINOs to 
high-dimensional PDEs that are relevant for the 
modeling of complex phenomena that demand subgrid scale 
precision, and which typically lead to computationally 
expensive simulations. We will also focus on developing 
methods that handle shocks effectively, since these 
phenomena are common in fluid dynamics and 
relativistic astrophysics, to mention a few. We will 
also continue our research 
program combining scientific visualization and 
accelerated 
computing to gain insights about what PINOs learn from 
data, and how this information may be used to enhance 
their performance when applied to experimental datasets~\cite{2017arXiv170208608D,2022arXiv220207399S,carvalho2019machine}.

It is our expectation that our 
AI surrogates may be used to replace computationally 
demanding numerical methods to learn 
PDEs in scientific software 
used to model multi-scale and multi-physics 
phenomena---e.g., numerical relativity simulations 
of gravitational wave 
sources, weather forecasting, etc.,---and eventually 
provide data-driven and physics informed solutions that 
more accurately describe and identify novel 
features and patterns in experimental data. 

\section*{Acknowledgments}
\noindent  This material is based upon work supported 
by Laboratory Directed Research and Development (LDRD) 
funding from Argonne National Laboratory, 
provided by the Director, Office of Science, of the 
U.S. Department of Energy under 
Contract No. DE-AC02-06CH11357. This research used 
resources of the Argonne 
Leadership Computing Facility, which is a DOE Office of 
Science User Facility supported under Contract 
DE-AC02-06CH11357. S.R. and E.A.H. gratefully 
acknowledge National Science Foundation 
award OAC-1931561. This work used the Extreme Science 
and Engineering Discovery Environment (XSEDE), which is 
supported by National Science Foundation grant 
number ACI-1548562. This work used the Extreme Science 
and Engineering Discovery Environment (XSEDE) 
Bridges-2 at the Pittsburgh Supercomputing Center 
through allocation TG-PHY160053. This research used 
the Delta advanced computing and data resource which 
is supported by the National Science Foundation 
(award OAC-2005572) and the State of Illinois. 
Delta is a joint effort of the University of 
Illinois Urbana-Champaign and its National 
Center for Supercomputing Applications. 

\bibliography{book_references}

\end{document}